\documentclass{aa}
\usepackage[varg]{txfonts}
\usepackage{graphicx}
\usepackage[version=3]{mhchem} 
\usepackage{longtable}
\bibpunct{(}{)}{;}{a}{}{,} 
\usepackage{color}

\begin{document}
	\newcommand{\VLA}{VLA~1623$-$2417}
	\newcommand{\vla}{VLA~1623}
	\newcommand{\vlaa}{VLA~1623 A}
	\newcommand{\vlab}{VLA~1623 B}
	\newcommand{\vlaw}{VLA~1623 W}
	
	\title{Revised SED of the triple protostellar system \VLA}
	
	\author{N. M. Murillo\inst{1} \and D. Harsono\inst{1} \and M. McClure\inst{2} \and S.-P. Lai\inst{3, 4} \and M. R. Hogerheijde\inst{1,2}}
	\institute{Leiden Observatory, Leiden University, P.O. Box 9513, 2300 RA, Leiden, the Netherlands \\ \email{nmurillo@strw.leidenuniv.nl}
	\and Anton Pannekoek Institute for Astronomy, University of Amsterdam, Science Park 904, 1098 XH, Amsterdam, The Netherlands
	\and Institute of Astronomy and Department of Physics, National Tsing Hua University, 101 Section 2 Kuang Fu Road, Hsinchu 30013, Taiwan
	\and Academia Sinica Institute of Astronomy and Astrophysics, P.O. Box 23-141, Taipei 10617, Taiwan}
	
	\abstract
	{\VLA~is a triple protostellar system deeply embedded in Ophiuchus A. Sources A and B have a separation of 1.1$\arcsec$, making their study difficult beyond the submillimeter regime. Lack of circumstellar gas emission suggested that \vlab~has a very cold envelope and is much younger than source A, generally considered the prototypical Class 0 source.}
	{We explore the consequences of new ALMA Band 9 data on the spectral energy distribution (SED) of \VLA~and their inferred nature.}
	{Using dust continuum observations spanning from centimeter to near-infrared wavelengths, the SED of each component in \VLA~is constructed and analysed.}
	{The ALMA Band 9 data presented here show that the SED of \VLA~B does not peak at 850$\mu$m as previously expected, but instead presents the same shape as \VLA~A at wavelengths shorter than 450 $\mu$m.}
	{The results presented here indicate that the previous assumption that the flux in \textit{Herschel} and Spitzer observations is solely dominated by \VLA~A is not valid, and instead, \VLA~B most likely contributes a significant fraction of the flux at $\lambda~<$ 450 $\mu$m. These results, however, do not explain the lack of circumstellar gas emission and puzzling nature of \VLA~B.}

	
	\titlerunning{Revised SED of \VLA}
	\authorrunning{Murillo et al.}
	
	\maketitle
	
	\section{Introduction}
	\label{sec:intro}
	\VLA~(hereafter \vla) is a deeply embedded protostellar system in Ophiuchus A (L1688), 
	and first observed by \citet{andre1990}.
	Recent observations place L1688 and \vla~ at a distance of 137.3 $\pm$ 1.2 pc \citep{ortiz2017}.
	\vla~is generally considered to be the prototypical Class 0 source \citep{andre1993}. 
	Observations at centimeter wavelengths with the Very Large Array (VLA) revealed several continuum peaks \citep{bontemps1997}, which were interpreted as the central object of \vla~and the jet knots A, B and C.
	However, observations of the outflow suggested more than one component is present in \vla~\citep{dent1995,yu1997,caratti2006}.
	Interferometric observations in submillimeter revealed that \vla~is actually a triple protostellar system \citep{looney2000,murillo2013a}, with each component driving an outflow of its own \citep{murillo2013a,santangelo2015}.
	The VLA interferometric centimeter peaks A and B matched with sources B and W, respectively \citep{ward2011,murillo2013a}. No submillimeter counterpart was observed for the centimeter peak C detected with the VLA.
	
	\vlaa~is brightest in the submillimeter wavelength range, presenting a centrally peaked source with extended dust emission coinciding with the Keplerian disk traced in \ce{C^{18}O} \citep{murillo2013b}. 
	\vlab~is separated from source A by 1.1$\arcsec$ to the west, and presents a strongly collimated outflow \citep{santangelo2015}. Interferometric observations do not detect warm (\ce{C^{18}O}, \ce{c-C3H2}) or cold (\ce{DCO+}, \ce{N2D+}, \ce{N2H+}) gas emission associated with \vlab~\citep{murillo2013b,murillo2016,murillo2018}. Submillimeter Array (SMA, \citealt{ho2004}) observations detected \ce{SO} peaking between sources A and B \citep{murillo2013a}. Even more interesting is the detection of a variable \ce{H2O} maser centered on \vlab~\citep{furuya2003}, detected between 8 to 13 km~s$^{-1}$. 
	Suggestions have been made that \vlab~could be a background object, such as a galaxy.
	The velocity shift of the \ce{H2O} maser would suggest a redshift of 3--4$\times$10$^{-5}$. Thus, if the maser is associated with \vlab, then this source is located within our galaxy.
	\vlaw, separated by 10$\arcsec$ from \vlaa, presents hints of a rotating envelope or disk-like structure, and a slightly different systemic velocity ($\sim$1 km~s$^{-1}$) from its companions \vlaa~and B ($\sim$3.5 km~s$^{-1}$) \citep{murillo2013b}.
	
	In this paper, we present Atacama Large Millimeter/submillimeter Array (ALMA) Cycle 2 Band 9 dust continuum observations of \vla. These data, combined with previous ALMA Band 6 and 7 observations, continuum fluxes found in literature, and fluxes from Spitzer and \textit{Herschel} observations, are used to construct and analyze the spectral energy distribution (SED) of each component in \vla.
	In addition, newly reduced Spitzer IRS spectra for \vla~is also presented here.
	
	\section{Observations}
	\label{sec:obs}
	
	Table~\ref{tab:alma} details the ALMA observations presented here, including the field of view, angular resolution and typical noise level. 

	Cycle 2 observations in Band 9 (Project code: 2012.1.00707.S; PI: D. Salter) have phase center $\alpha_{J2000}$~=~16:26:26.42; $\delta_{J2000}$~=~$-$24:24:30.0. However the observations required rephasing, and the phase center was shifted to the position of \vlaa, $\alpha_{J2000}$~=~16:26:26.39; $\delta_{J2000}$~=~$-$24:24:30.688, consistent with the Cycle 2 Band 6 and 7 observations (see below). The spectral set-up was configured to observe mainly \ce{CO}, \ce{HCO+} and continuum. J1924$-$2914 was used for bandpass calibration. Phase calibration was done using the check source J1700-2610, with a time bin of 24 seconds, instead of 6 seconds. Flux calibration was done with J1733$-$130. Ceres was observed, but it was resolved on some baselines, and thus did not provide a reliable flux calibration. Observations toward \vla~did not have any system temperature measurements, so the system temperature of the phase calibrators was used.

	Cycle 2 observations in Bands 6 and 7 (Project code: 2013.0.01004; PI: S.-P. Lai) have a phase center of $\alpha_{J2000}$~=~16:26:26.390; $\delta_{J2000}$~=~$-$24:24:30.688, and targeted several molecular lines 
	(e.g., \ce{DCO+}, \ce{CO} isotopologues and \ce{c-C3H2} in Band 6, \ce{DCO+}, \ce{N2H+} and \ce{H2D+} in Band 7) 
	along with continuum. The molecular line observations are detailed in \citet{santangelo2015,murillo2015,murillo2018}.
	ALMA Cycle 0 Band 6 observations (Project code: 2011.0.00902.S; PI: N. Murillo) of \vla, with phase center $\alpha_{J2000}$~=~16:26:26.419; $\delta_{J2000}$~=~$-$24:24:29.988, slightly offset from the position of \vlaa, were previously reported in \citet{murillo2013b}.
	
	The Spitzer IRS spectra are new reductions of serendipitous observations of \vla. The \vlaw~and blended \vlaa~\& B components were aligned along the LL slit of the second target in the clustered AOR 12697600 (PID 2) and are visible in the second nod. The basic-calibrated data were downloaded from the Spitzer Heritage Archive and reduced by hand using the AdOpt optimal point source extraction routines from the CASSIS low-resolution pipeline update, version 7 \citep{lebouteiller2010} provided in SMART \citep{higdon2004}. The \vlaw~and blended \vlaa~\& B components were fit simulataneously, along with a low-order polynomial sky background, at each wavelength. The resulting spectra are complete in LL (14-35 microns) for source W, but the flux of the blended sources A \& B is only detected longwards of 19 microns.
	
	\begin{table*}
		\centering
		\caption{ALMA observations}
		\begin{tabular}{c c c c c c c c c}
			\hline \hline
			Project ID & Band & Wavelength & UV-baseline range & Largest scale & Field of view & Beam & P.A. & Noise \\
			 & & & k$\lambda$ & $\arcsec$ & $\arcsec$ & $\arcsec$ & $^{\circ}$ & mJy~beam$^{-1}$ \\
			\hline
			 2011.0.00902.S & 6 & 1.3 $-$ 1.4 mm & 25 $\sim$ 310 & 2.5 & 28.1 & 0.79$\arcsec\times$0.54$\arcsec$ & $-$85.9 & 3.5 \\
			 2013.0.01004.S & 6 & 1.3 $-$ 1.4 mm & 18 $\sim$ 791 & 2.0 & 28.8 & 0.45$\arcsec\times$0.27$\arcsec$ & 78.4 & 12 \\
			 2013.0.01004.S & 7 & 828 $-$ 832 $\mu$m & 18 $\sim$ 420 & 3.1 & 17.2 & 0.86$\arcsec\times$0.53$\arcsec$ & $-$87.1 & 4.2 \\
			 2012.1.00707.S & 9 & 421 $-$ 429 $\mu$m & 30 $\sim$ 2280 & 1.6 & 9.0 & 0.25$\arcsec\times$0.14$\arcsec$\tablefootmark{a} & $-$85.1 & 23.9\tablefootmark{a} \\
			 2012.1.00707.S & 9 & 421 $-$ 429 $\mu$m & 30 $\sim$ 2280 & 1.6 & 9.0 & 0.17$\arcsec\times$0.1$\arcsec$\tablefootmark{b} & 85.6 & 14.3\tablefootmark{b} \\
			\hline
		\end{tabular}
	\tablefoot{
		\tablefoottext{a}{Continuum image generated with natural weighting.}
		\tablefoottext{b}{Continuum image generated with robust (Briggs) weighting.}
	}
	\label{tab:alma}
	\end{table*}

	\begin{figure}
		\includegraphics[width=\columnwidth]{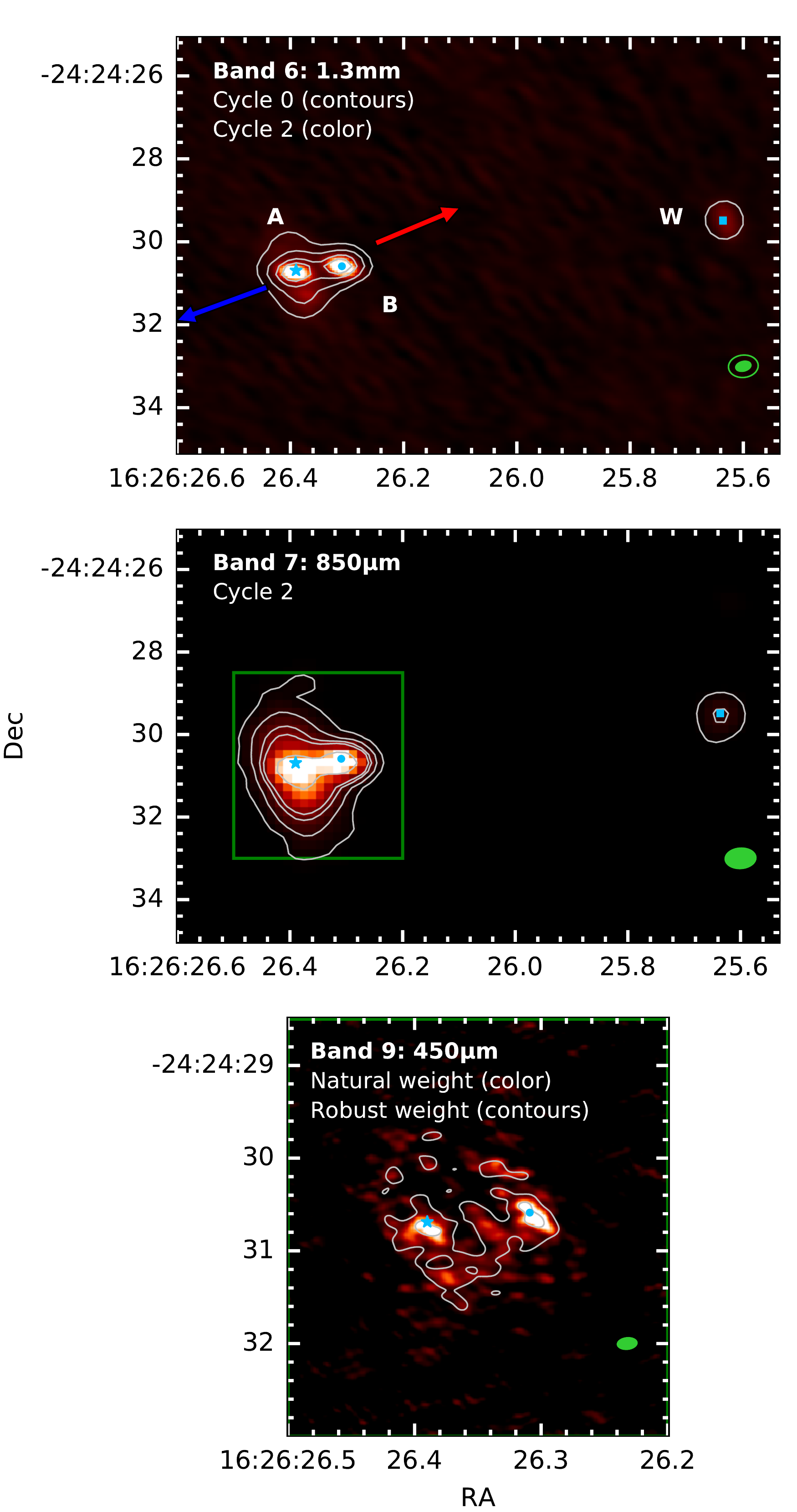} 
		\caption{\vla~observed with ALMA. The phase center for all observations is similar and located around the position of \vlaa. Contours are in steps of 3, 8, 15, 20, 50$\sigma$, with $\sigma$ listed in Table~\ref{tab:alma}. The green ellipses show the beamsize of the observations. The green box in the middle panel outlines the area of the Band 9 observations presented in the bottom panel.}
		\label{fig:cont}
	\end{figure}
	
	\section{Results}
	\label{sec:res}
	
	Measured fluxes from ALMA observations are listed in Table~\ref{tab:sed}.
	\vlaa~presents a central bright component surrounded by extended emission in all bands (Fig.~\ref{fig:cont}). In the Band 9 observations, source A does not show further multiplicity, and the clumpiness of the extended dust emission is most likely due to the difficulty of recovering extended structure with high spatial resolution observations. The extended dust emission from Cycle 2 Band 6 data was found to have a radius of $\sim$100 AU \citep{persson2016}, smaller than that of the Keplerian disk (150 AU) traced with \ce{C^{18}O} \citep{murillo2013b}.
	In contrast, \vlab~is observed to be compact, almost point-like in all Bands. The extended emission in the Band 9 image is most likely not real, and product of the difficulty of recovering structure with high spatial resolution observations. Similarly, \vlaw~is also point-like in Bands 6 and 7, but with a continuum flux 3 to 5 times weaker than \vlab~(Table~\ref{tab:sed}). 
	Fluxes for each component are measured using a region which is defined by the 3$\sigma$ contour of the continuum emission. For Band 9 continuum, we use the same region used for Band 6. For \vlaa, the extended dust disk component is included in the region used for flux measurement. Measured fluxes are listed in Table~\ref{tab:sed}. \vlaa~is brighter than \vlab~by a factor up to 2, although this difference in fluxes might be caused by the inclusion of the dust disk in the flux of \vlaa, and assuming that \vlab~has no disk. \vlaw~is dimmer than \vlaa~and B by about a factor of 5 to 12.
	
	\begin{figure}
	\includegraphics[width=\columnwidth]{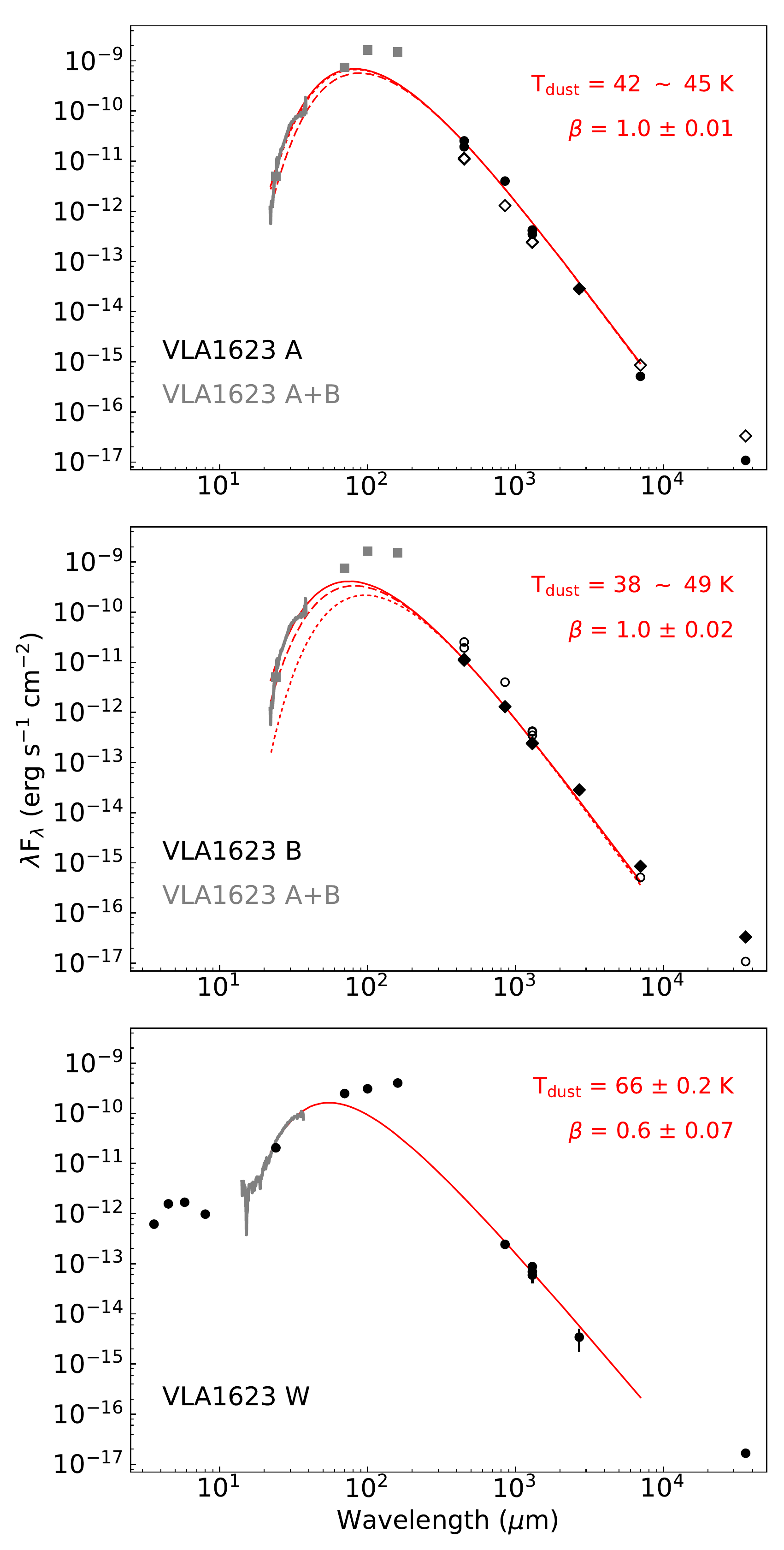} 
	\caption{SED of \vla~including the Spitzer IRS spectra. The red curves show the best fit graybody to each SED, with the dust temperature $T_{\rm dust}$ and dust emissivity index $\beta$ indicated in each panel. For \vlaa~(top) and B (middle), the dashed curves show the best fit assuming that the emission at $\lambda <$450 $\mu$m is evenly distributed among the two sources. The dotted curve assumes \vlab~only contributes one-tenth of the total emission for $\lambda <$450 $\mu$m. The open symbols in the top and middle panel are the fluxes for \vlab~and \vlaa, respectively, and are included for ease of comparison.}
	\label{fig:SED}
	\end{figure}
	
	\section{Analysis}
	\label{sec:analysis}
	
	In order to construct the SEDs of \vla, the fluxes presented in this work are used, together with those found in literature. 
	To constrain the peak of the SEDs, archival \textit{Herschel Space Observatory} \citep{pilbratt2010} PACS photometric maps for Ophiuchus from the Gould Belt Survey \citep{andre2010,pezzuto2012} and from the program of PI: P. {\'A}brah{\'a}m \citep{alves2013} were obtained from the Herschel Science Archive (Observation IDs: 134222148 and 1342238816). The maps produced with the JScanmap method were used.
	The 160 and 100 $\mu$m maps show the emission toward \vla~to be concentrated mainly on the position of \vlaa~and B. A distinct \vlaw~emission is not resolved at either wavelength, most likely due to the resolution of the \textit{Herschel} observations.
	At 70 $\mu$m, two distinct continuum emission components are detected in the maps, one centered on \vlaa~\& B, and the other on \vlaw. Both components show extended emission, most likely arising from their outflow cavities.
	Due to the nature of the sources, PSF photometry is performed on the PACS maps to extract fluxes using \textit{StarFinder} \citep{diolaiti2000}.
	The PSF for each wavelength is extracted from the corresponding PACS maps using sources that appear as point sources and have little to no extended emission around them. An aperture mask is applied to the PSF, thus minimizing the extented emission that can be included during photometry. Due to the aperture mask, an aperture correction factor obtained from \citet{balog2014} is applied to the fluxes. At 160 and 100 $\mu$m, deblending of the detected sources was done in order to obtain fluxes for \vlaw. The method is further detailed in \citet{murillo2016} and \citet{diolaiti2000}. 

	Table~\ref{tab:sed} lists the fluxes and Fig.~\ref{fig:SED} shows the SED of each component. At wavelengths smaller than 450 $\mu$m, \vlaa~and B are spatially unresolved.
	Observations with SMA at 1.3 mm and BIMA at 2.7 mm \citep{looney2000,murillo2013a} detect all three components. In both datasets, \vlaa~is the brightest of the three sources, while \vlab~and W are point-like. The extended emission around \vlaa~is resolved with the SMA (synthesized beam: 0.61$\arcsec~\times$ 0.43$\arcsec$, 10$^{\circ}$) but not in the BIMA observations (synthesized beam: 0.95$\arcsec~\times$ 0.39$\arcsec$, 18$^{\circ}$).
    Spitzer observations detect \vlaw~in all IRAC and MIPS channels, while unresolved \vlaa~and B are detected only in MIPS1 \citep{murillo2013a}, indicating that both sources are deeply embedded. IRS spectra for \vlaw, and unresolved \vlaa~and B are included in the SEDs.
	
	The SEDs are fit using the Scipy method curve$\_$fit\footnote{https://www.scipy.org/} with a single-temperature graybody, and with optical depth $\tau\propto\nu^{\beta}$, where $\beta$ is the dust emissivity index. The fit is done in the wavelength range from 7 mm to 22 $\mu$m. The best fit curves are shown in Fig.~\ref{fig:SED}. The graybody fits are practically the same with or without the \textit{Herschel} PACS fluxes, which is to be expected since the \textit{Herschel} fluxes are picking up the envelope material as well.
	In the submillimeter regime, the SEDs of \vlaa~and B are quite similar, as shown in Fig.~\ref{fig:SED} with open and filled symbols.
	For sources A and B, the unresolved emission at $\lambda <$ 450$\mu$m is fitted in three ways. First, assuming all the infrared emission belongs to one of the sources (Fig.~\ref{fig:SED} solid red line). Second, the infrared emission is evenly distributed between both sources (Fig.~\ref{fig:SED} dashed red line). Third, \vlab~only contributes one-tenth of the infrared emission (Fig.~\ref{fig:SED} dotted red line). The resulting dust temperatures $T_{\rm dust}$ are similar for all three fits, and for both systems, between 40 to 50 K. However, it is not certain if the emission is evenly distributed at all wavelengths or varies with wavelength. Thus, there are large uncertainties in the SED fit of \vlaa~and B. However, the derived dust temperature for \vlab~(38 to 49 K) is larger by an order of magnitude than previously expected ($T_{\rm dust}$ = 5 K, \citealt{murillo2013a}).
	The derived dust temperature for \vlaw~(66 K) is also higher than previously reported ($T_{\rm dust}$ = 41 K, \citealt{murillo2013a}). The higher dust temperature is most likely due to the addition of the IRS spectra to the SED, which was not included in previous studies.
	
	The bolometric luminosity is calculated based on the curve of the fits to the constructed SEDs using the trapezoid method.
	Considering the three cases used for SED fitting, the bolometric luminosity of \vlaa~is between $\sim$0.4 -- 0.5 L$_{\odot}$, while for \vlab~it is between $\sim $0.2 -- 0.3 L$_{\odot}$. It must be stressed that due to the SED being unresolved at wavelengths shorter than 450 $\mu$m for \vlaa~and B, parameters derived from the SED are uncertain, and these bolometric luminosities are rough estimates. In addition, source A appears brighter since the emission from the dust disk is being included. For \vlaw, the derived bolometric luminosity is on the order of 0.13 $\pm$ 0.01 L$_{\odot}$.

	To calculate the envelope mass, the formula from \citet{jorgensen2009}, which takes into account that brighter sources have somewhat higher dust temperatures, is used and expressed as:
	\begin{equation}
	M_{\rm env} = 0.44 M_{\odot}~\left(\frac{L_{\rm bol}}{1~L_{\odot}}\right)^{-0.36}~\left(\frac{S_{850\mu m}}{1~{\rm Jy~beam}^{-1}}\right)^{1.2}~\left(\frac{d}{125~{\rm pc}}\right)^{1.2}.
	\end{equation}
	For this calculation, the 850 $\mu$m fluxes obtained with ALMA Cycle 2 Band 7 observations are used, in conjunction with the bolometric luminosities derived from the SEDs and described above.
	This method yields a mass of $\sim$0.8 and 0.2 M$_{\odot}$ for \vlaa~and B, respectively, for a distance of 137.3 pc. For \vlaw, the mass is 0.04 $\pm$ 0.005 M$_{\odot}$. If using the previous distance to \vla~of 120 pc \citep{loinard2008}, the masses are decreased by 15\%.
	Several caveats need to be highlighted here. The 850 $\mu$m flux from ALMA Cycle 2 observations only recovers emission from the disk and the inner envelope (up to 3$\arcsec$, $\sim$400 AU), but not from the large scale gas ($>$ 400 AU). Thus the mass estimate reported here is not representative of the whole envelope mass. Furthermore, if \vlab~is observed through the envelope of A, then the mass estimate for B is in fact not indicative of its actual mass.

	\begin{table*}
		\centering
		\caption{SED for each component in VLA1623} 
		\begin{tabular}{c c c c c c c c c}
			\hline \hline
			 & \multicolumn{2}{c}{VLA~1623 A} & \multicolumn{2}{c}{VLA~1623 B} & \multicolumn{2}{c}{VLA~1623 W} & & \\
			\hline
			RA & \multicolumn{2}{c}{16:26:26.390} & \multicolumn{2}{c}{16:26:26.309} & \multicolumn{2}{c}{16:26:25.636} & & \\
			Dec & \multicolumn{2}{c}{$-$24:24:30.688} & \multicolumn{2}{c}{$-$24:24:30.588} & \multicolumn{2}{c}{$-$24:24:29.488} & & \\
			\hline
			Wavelength & S$_{\rm int}$ & error & S$_{\rm int}$ & error & S$_{\rm int}$ & error & Ref. & Telescope \\
			$\mu$m & mJy & mJy & mJy & mJy & mJy & mJy & & \\
			\hline
			3.6 &  &  &  &  & 0.74 & 0.12 & 1 & Spitzer/IRAC1 \\
			4.5 &  &  &  &  & 2.35 & 0.19 & 1 & Spitzer/IRAC2\\
			5.8 &  &  &  &  & 3.26 & 0.33 & 1 & Spitzer/IRAC3\\
			8.0 &  &  &  &  & 2.60 & 0.26 & 1 & Spitzer/IRAC4\\
			24 &  \multicolumn{4}{c}{40.4 $\pm$ 4.2\tablefootmark{a}}   & 164.26 & 4.54 & 2,1 & Spitzer/MIPS1\\
			70\tablefootmark{b} &  \multicolumn{4}{c}{17424.5 $\pm$ 3.2\tablefootmark{a}}   & 5790.8 & 3.88 & 3 & Herschel/PACS \\
			100\tablefootmark{b} &  \multicolumn{4}{c}{55092.3 $\pm$ 8.7\tablefootmark{a}}   & 10284.9\tablefootmark{c} & 8.6 & 3 & Herschel/PACS \\
			160\tablefootmark{b} &  \multicolumn{4}{c}{81340.6 $\pm$ 10.3\tablefootmark{a}}   & 21457.2\tablefootmark{c} & 11.5 & 3 & Herschel/PACS \\
			450 & 3820.0 & 4.2 & 1723.0 & 5.7 &  & & 3 & ALMA/Band 9 Natural weighting \\
			450 & 2910.0 & 6.4 & 1653.0 & 6.8 &  & & 3 & ALMA/Band 9 Briggs weighting\\
			850 & 1139.5 & 3.4 & 373.0 & 4.6 & 69.0 & 7.0 & 3 & ALMA/Band 7 \\
			1300 & 184.0 & 13.8 & 104.0 & 15.3 & 38.0 & 3.8 & 3 & ALMA/Band 6 Cycle 0 \\
			1300 & 178.0 & 5.1 & 106.0 & 5.3 & 30.0 & 3.0 & 3 & ALMA/Band 6 Cycle 2 \\
			1300 & 152.0 & 8.0 & 106.0 & 8.0 & 25.7 & 8.0 & 4 & SMA \\
			2700 & 25.5 & 6.3 & 25.8 & 3.5 & 3.1 & 1.5 & 5 & BIMA \\
			7000 & 1.2 & 0.2 & 2 & 0.2 &  & & 6 & VLA \\
			36000 & 0.13 & 0.02 & 0.4 & 0.02 & 0.2 & 0.02 & 6,7 & VLA \\
			44000 &  \multicolumn{2}{c}{$<$0.2}  &  \multicolumn{2}{c}{$<$0.2}  &  & & 6 & MERLIN \\
			\hline
		\end{tabular}
		\tablefoot{
			\tablefoottext{a}{Unresolved fluxes.}
			\tablefoottext{b}{Fluxes obtained with \textit{StarFinder} PSF photometry using an aperture mask of 11.2$\arcsec$, 12.8$\arcsec$ and 19.2$\arcsec$ for 70, 100 and 160 $\mu$m, respectively. The aperture correction factors are 0.791, 0.795 and 0.789 for 70, 100 and 160 $\mu$m, respectively \citep{balog2014}.}
			\tablefoottext{c}{\vlaw~fluxes obtained using the deblend option in \textit{StarFinder}, since no distinct peak for \vlaw~is observed.}
		}
		\tablebib{(1) \citet{gutermuth2009}; (2) c2d catalog; (3) this work; (4) \citet{murillo2013a}; (5) \citet{looney2000}; (6) \citet{ward2011}; (7) \citet{bontemps1997}}
		\label{tab:sed}
	\end{table*}
	
	\section{Summary}
	\label{sec:conc.}
	
	In this paper we present new ALMA Band 9 continuum data toward \vla, resolving the components A and B. PSF photometry of \textit{Herschel} PACS maps provides unresolved fluxes between 70 to 160 $\mu$m, which trace the peak of the SED.
	The updated SEDs show that \vlab~does not peak at 1000 $\mu$m, as suggested by \citet{murillo2013a} based on SMA 850 $\mu$m observations, but might instead peak at around 100$\mu$m, similar to \vlaa.
	The best fit using a graybody indicates that the dust temperature of both sources is in the 40 to 50 K range. There are large uncertainties, however, due to the unresolved fluxes of \vlaa~and B at wavelengths shorter than 450 $\mu$m, and thus the derived dust temperature should be considered with caution.
	With these results, the previous assumption that the continuum flux at shorter wavelengths is dominated by \vlaa~is not well based, and instead \vlab~could be the dominant source in the infrared regime.
	
	Deriving source parameters or evolutionary stage from unresolved SEDs is non-trivial, and can lead to under- or over-estimation of the parameters (see e.g., \citealt{murillo2016}). Keeping this caveat in mind, the bolometric luminosity derived for \vlaa~and B from the SEDs presented in this work, together with the rough estimate of the disk and inner envelope mass on scales up to 400 AU, provide constraints on the source parameters for each component, albeit with large uncertainties. 
	The bolometric luminosity for \vlaa~and B varies by a factor of $\sim$2, and the mass of the disk plus inner envelope varies by a factor of $\sim$4 since it is dependent on the luminosity as well as the 850 $\mu$m flux. All parameters derived from the SED indicate that both sources are Class 0. In the scenario where source B would dominate the infrared emission, its bolometric luminosity would rise and the mass would decrease.
	For \vlaw, the luminosity is lower by a factor of $\sim$4 compared to source A, while the envelope mass is lower by a factor of 20 compared to that of source A, consistent with its classification as a Class I source.
	
	Along with separating the fluxes of components in a multiple protostellar system, the variability of the sources must also be taken into account when understanding the nature of these systems. Recent studies show variation in the brightness of protostellar systems, possibly due to episodic accretion processes (e.g., \citealt{herczeg2017,johnstone2018}). 
	For the case of \vla, however, all three components have not shown significant variations in brightness in the submillimeter range. This can be noted in the 1.3 mm fluxes (Table~\ref{tab:sed}) which have been obtained with the SMA (2007 and 2009), and with ALMA in Cycle 0 (2011) and Cycle 2 (2013). 
	The relative similarity of the fluxes at 1.3 mm suggests that there has been no large variations in brightness. There could be variations on the order of 10\% at 1.3 mm, but given the uncertainties of the flux calibration, such variations would not be detected.
	In addition to variability in brightness, chemical signatures in the envelope gas can be used to probe episodic accretion \citep{visser2015,jorgensen2015,frimann2017,hsieh2018}.
	Considering the characterized gas structure of \vla~\citep{murillo2013b,murillo2015,murillo2018}, no chemical signatures of luminosity variations toward the components of \vla~are detected.
	
	The presented results contribute to the puzzle of \vlab. The derived dust temperature from the SED fitting (40 to 50 K) suggests there should be warm gas emission from this source. However, no circumstellar gas emission in either cold or warm molecules \citep{murillo2013b,murillo2015,murillo2018} has been detected to be associated with \vlab. In contrast, \vlab~drives a fast and collimated jet and outflow \citep{santangelo2015}, and presents a variable \ce{H2O} maser \citep{furuya2003}.
	The lack of detected circumstellar gas emission around \vlab~is difficult to explain.
	One possibility could be gas stripping due to interaction with the outflow from \vlaa~or \vlaw~(e.g. B1-b: \citealt{hirano2014}), but this would not explain why the dust was not stripped or presence of the outflow.
	Another explanation would be that \vlab~is more evolved than \vlaa. However this would not fully explain the complete lack of gas in submillimeter when \vlaw~does present circumstellar gas \citep{murillo2013b}.
	A final possibility would be that \vlab~is being obscured, either by the envelope of \vlaa, or circumstellar matter with high optical depth. This would also introduce further uncertainties to the calculated envelope mass.
	Observations with subarcsecond resolution in the infrared are required to determine the nature of this object, for which JWST will be ideal.

\begin{acknowledgements}
	This paper made use of the following ALMA data: ADS/JAO.ALMA 2011.0.00902.S, 2012.1.00707.S, and 2013.1.01004.S. ALMA is a partnership of ESO (representing its member states), NSF (USA), and NINS (Japan), together with NRC (Canada) and NSC and ASIAA (Taiwan), in cooperation with the Republic of Chile. The Joint ALMA Observatory is operated by ESO, AUI/NRAO, and NAOJ.
	The authors acknowledge assistance from Allegro, the European ALMA Regional Center node in the Netherlands.
	Herschel is an ESA space observatory with science instruments provided by European-led Principal Investigator consortia and with important participation from NASA.
	PACS has been developed by a consortium of institutes led by MPE (Germany) and including UVIE (Austria); KU Leuven, CSL, IMEC (Belgium); CEA, LAM (France); MPIA (Germany); INAF-IFSI/OAA/OAP/OAT, LENS, SISSA (Italy); IAC (Spain). This development has been supported by the funding agencies BMVIT (Austria), ESA-PRODEX (Belgium), CEA/CNES (France), DLR (Germany), ASI/INAF (Italy), and CICYT/MCYT (Spain).
	This work is based in part on observations made with the Spitzer Space Telescope, which is operated by the Jet Propulsion Laboratory, California Institute of Technology under a contract with NASA.
	NMM and DH are supported by the European Union A-ERC grant 291141 CHEMPLAN, by the Netherlands Research School for Astronomy (NOVA), by a Royal Netherlands Academy of Arts and Sciences (KNAW) professor prize.
	MM receives funding through a Marie Sk\l{}odowska-Curie action in the Horizon 2020 program.
	SPL acknowledges support from the Ministry of Science and Technology of Taiwan with Grant MOST 106-2119-M-007-021-MY3.
	MRH is support by grant 614.001.352 from the Netherlands Organisation for Scientific Research (NWO).
	The authors wish to thank Doug Johnstone for the helpful and insightful comments that greatly improved this work.
\end{acknowledgements}

\bibliographystyle{aa}
\bibliography{vla1623_letter.bib}

\end{document}